\documentclass[11pt]{article}
\usepackage{latexsym}
\usepackage{amsmath} 
\usepackage{amscd}     \usepackage{amsxtra}
\usepackage{upref}     \usepackage{amsthm}
\usepackage{amssymb}   
\usepackage{amsfonts}
\begin{document}

\title{{\bf Complex Structures in Electrodynamics}}

\author{{\bf Stoil Donev}\footnote{e-mail:
 sdonev@inrne.bas.bg} \\ Institute for Nuclear Research and Nuclear
Energy,\\ Bulg.Acad.Sci., 1784 Sofia, blvd.Tzarigradsko chaussee 72\\
Bulgaria\\}

\date{}
\maketitle

\begin{abstract}
In this paper we show that the basic external (i.e. not determined by the
equations) object in Classical electrodynamics equations is a {\it complex
structure}. In the 3-dimensional standard form of Maxwell equations this
complex structure $\mathcal{I}$ participates implicitly in the equations and
its presence is responsible for the so called {\it duality invariance}.  We
give a new form of the equations showing explicitly the participation of
$\mathcal{I}$. In the 4-dimensional formulation the complex structure is
extracted directly from the equations, it appears as a linear map $\Phi$ in
the space of 2-forms on $\mathbb{R}^4$.  It is shown also that $\Phi$ may
appear through the equivariance properties of the new formulation of the
theory. Further we show how this complex structure $\Phi$ combines with the
Poincar\'e isomorphism $\mathfrak{P}$ between the 2-forms and 2-tensors to
generate all well known and used in the theory (pseudo)metric constructions
on $\mathbb{R}^4$, and to define the conformal symmetry properties. The
equations of Extended Electrodynamics (EED) do not also need these
pseudometrics as beforehand necessary structures.  A new formulation of the
EED equations in terms of a generalized Lie derivative is given.

\end{abstract}

\section{Introduction}

We begin with two examples, showing that meeting with implicitly
participating objects in some equations of mathematical physics is not an
unknown phenomenon. Recall the wave D'Alembert equation (in standard form)
$$
U_{tt}-c^2\left(U_{xx}+U_{yy}+U_{zz}\right)=0.
$$
Except the constant $c$, no external objects participate in this equation.
During the first half of 20th century a new understanding of this equation
was created, namely, that a new external object participates implicitly in it
and it is the pseudeuclidean metric tensor $g^{\mu\nu}$,
$-g^{11}=-g^{22}=-g^{33}=g^{44}=1$ on $\mathbb{R}^4$, so that the true form
of this equation should read

$$
g^{\mu\nu}\frac{\partial^2 U}{\partial x^\mu \partial x^\nu} =0 .
$$
This form of the equation has general covariance, i.e. the coordinates used
may be arbitrary. The symmetries of the equation, coming from
transformations of the base manifold $\mathbb{R}^4$, as well as many other
of its important properties, seem to be determined by the
isometries of $g^{\mu\nu}$.

The later on studies brought another view,
saying that the Hodge $*$-operator, defined by $g_{\mu\nu}$, is the essential
object, and the equation acquired any of the two {\it coordinate free} forms

$$
\mathbf{d}*\mathbf{d}U=0, \quad (\delta\mathbf{d}+\mathbf{d}\delta)U=0
$$
where $\mathbf{d}$ is the exterior derivative, and
$\delta=(-1)^p*^{-1}\mathbf{d}*$ is the coderivative ($\delta U\equiv 0$).

If we continue this
process of a {\it precise revealing the structures}, defining the equation,
we would come to the conclusion that, in fact, the pseudometric tensor
$g_{\mu\nu}$ is {\it not} needed, and the {\it necessary and sufficient
external} structure needed to give a coordinate free form of the equation is a
linear map $f:\Lambda^1(\mathbb{R}^4)\to \Lambda^3(\mathbb{R}^4)$ (in fact, a
linear isomorphism) from the space of 1-forms on $\mathbb{R}^4$ to the space
of 3-forms on $\mathbb{R}^4$ defined (in canonical coordinates) by
\begin{align*}
& f(dx)=-dy\wedge dz\wedge d\xi, \ f(dy)=dx\wedge dz\wedge d\xi,         \\
& f(dz)=-dx\wedge dy\wedge d\xi, \ f(d\xi)=-dx\wedge dy\wedge dz,
\end{align*}
where $\xi=ct$. Then the above wave equation will read (in canonical
coordinates)

$$
\mathbf{d}f(\mathbf{d}U)=
\left(-U_{xx}-U_{yy}-U_{zz}+U_{\xi\xi}\right)
dx\wedge dy\wedge dz\wedge d\xi=0.
$$
That the linear map $f$ can be defined by the pseudometric $g$ through the
Hodge $*$ is obvious, but it is also evident that $f$ can be defined
independently of $g$. Hence, since the exterior derivative $\mathbf{d}$ is
defined only by the differential structure on $\mathbb{R}^4$, the symmetries
and the other properties of the D'Alembert wave equation are determined
entirely by $f$.

Another example comes from mechanics in its hamiltonian formulation. If $q^i$
and $p^i$ are the classical coordinates and momentum components and $H$ is
the hamiltonian, then we have the well known hamiltonian form of the basic
equations of classical mechanics

$$
\dot{q^i}=\frac{\partial H}{\partial p^i},\
\dot{p^i}=-\frac{\partial H}{\partial q^i}.
$$
These equations, have been studied quite a long time (nearly a century)
before to become clear that there is an external object implicitly
participating in these equations, namely the {\it symplectic} 2-form
$\omega=dq^i\wedge dp^i$ on the cotangent bundle of $\mathbb{R}^3$. It became
clear then that the symmetries of the equations, called usually {\it
canonical transformations}, coincide with the symmetries of the symplectic
2-form $\omega$. This fact is of so great importance in mechanics that could
hardly be overestimated. The tremendous mathematical development (called
now {\it symplectic geometry}) that followed this simple observation
undoubtedly proves this and shows the importance of having a clear and full
knowledge of {\it every detail} when studying an equation.

Maxwell equations of Classical electrodynamics are of exclusive importance in
all theoretical and mathematical physics.  Their deep study at the end of
19th and the beginning of 20th centuries gave birth to the relativistic view
on the physical world. Their duality properties generated the modern
"dualities" in field (superstring) theories. Their gauge interpretation
brought modern gauge theory, and the leading role of gauge theory in today's
field theory is out of any doubt.  Therefore, revealing the right external
mathematical structures these equations use, we consider as a meaningful and
important task.

In this paper, following the ideas previously stated in [1,2], we shall show
that the standard complex structure $\mathcal{I}$ in $\mathbb{R}^2$
(here and further under {\it complex structure} we meen a linear map $\Phi$
having the property $\Phi_\circ \Phi=-id$) implicitly participates in the
3-dimensional formulation of Maxwell equations, making quite obvious their
dual symmetry properties.  The new mathematical representation of the field
through the $\mathbb{R}^2$ valued 1-form $\omega=\mathbf{E}\otimes
e_1+\mathbf{B}\otimes e_2$ on $\mathbb{R}^3$ and through the above mentioned
complex structure $\mathcal{I}$ we consider as a more adequate and a more
appropriate one for the revealing the symmetry properties of Maxwell
equations. In some sense we follow the symplectic mechanics development:  the
symplectic 2-form in hamilton equations interprets appropriately the minus
sign in the second group equations and the canonical transformations;
similarly, the complex structure $\mathcal{I}$ will interpret appropriately
the minus sign in one of the "curl" Maxwell equations and their dual
symmetry.

In the 4-dimensional differential form formulation of Maxwell equations
we get the possibility to transform the presence of the complex structure
$\mathcal{I}$ in the 3-d form of the equations to a presence of a special
complex structure $\Phi:\Lambda^2(\mathbb{R}^4)\to \Lambda^2(\mathbb{R}^4)$
in the space of differential 2-forms on $\mathbb{R}^4$.  And this linear map
$\Phi$ is, in fact, the necessary and sufficient external object that is
needed to build the whole picture.

We would like to specially emphasize that we aim to reveal the {\it precise
external structures} used by Maxwell equations: {\it no more,
and no less}. All further reinterpretations of these structures in
terms of other objects and operations we consider as a second step.
For example, we may define Maxwell
equations on $\mathbb{R}^4$ through the Hodge $*$-operator defined by the
pseudometric $g$: $\mathbf{d}F=0,\ \mathbf{d}*F=0$, but the important moment
is that the second equation $\mathbf{d}*F=0$ does {\it not} make use of the
$*$-operator, it uses {\it only one} of its properties, namely the
property that it defines a {\it concrete complex structure} in
$\Lambda^2(\mathbb{R}^4)$ and {\it nothing more}, and this complex structure
may be introduced {\it without making use of the pseudometric} $g$. So, if we
start from the equations, we have to try to build all needed structures in
terms of those already introduced by the very equations, and we should
introduce new ones {\it only if it is impossible} to define them through the
available ones. This is the philosophy we are going to follow in this paper.

\section{The Complex Structure in the Standard 3-d Formulation of Maxwell
Equations}

We consider the pure field Maxwell equations
\begin{equation}
{\rm curl}\,\mathbf{E}+
\frac 1c      \frac{\partial {\mathbf{B}}} {\partial t}=0
, \quad {\rm div}\,\mathbf{B}=0,             
\end{equation}
\begin{equation}
{\rm curl}\,\mathbf{B} -
	\frac 1c \frac{\partial {\mathbf{E}}} {\partial t}=0 ,
\quad {\rm div}\,\mathbf{E}=0.                                 
\end{equation}

First we note, that because of the linearity of these equations if
$({\mathbf{E}}_i,\mathbf{B}_i), i=1,2,...$  are a collection of solutions,
then every couple of linear combinations of the form
\begin{equation}
{\mathbf{E}}=a_i{\mathbf{E}}_i,\ \mathbf{B}=a_i\mathbf{B}_i   
\end{equation}
(sum over the repeated $i=1,2,...$) with arbitrary constants $(a_i)$
gives a new solution.

The important observation made by Heaviside [3], and later considered
by Larmor [4], is that the substitution
\begin{equation}
{\mathbf{E}}\rightarrow -\mathbf{B},\quad
\mathbf{B}\rightarrow {\mathbf{E}}      
\end{equation}
transforms the first couple (1) of the pure field Maxwell equations into the
second couple (2), and, vice versa, the second couple (2) is transformed into
the first one (1). This symmetry transformation (4) of the pure field Maxwell
equations is called {\it special duality transformation}, or
SD-transformation.  It clearly shows that the electric and magnetic
components of the pure electromagnetic field are interchangeable and the
interchange (4) transforms solution into solution. This feature of the pure
electromagnetic field reveals its {\it dual} nature.

It is important to note that the SD-transformation (4) does not
change the energy density $8\pi{\bf w}={\mathbf{E}}^2+\mathbf{B}^2$,
the Poynting vector $4\pi\mathbf{S}=c({\mathbf{E}}\times\mathbf{B})$ , and the
(nonlinear) Poynting relation
\[
\frac {\partial}{\partial t} \frac {{\mathbf{E}}^2+\mathbf{B}^2}{8\pi}=
-{\rm div}\,\mathbf{S}.
\]
Hence, from energy-momentum point of view two dual, in the sense of (4),
solutions are indistinguishable.

Note that the substitution (4) may be considered as a transformation of the
following kind:
\begin{equation}
(\mathbf{E},\mathbf{B})
	\begin{Vmatrix}
0   &1
\\
-1  &0
	\end{Vmatrix}
=
(-\mathbf{B},\mathbf{E}).
\end{equation}
The following question now arises naturally: do there exist constants
$(a,b,m,n)$, such that the linear combinations
\begin{equation}
{\mathbf{E}}'=a{\mathbf{E}}+m\mathbf{B},\
\mathbf{B}'=b{\mathbf{E}}+n\mathbf{B},       
\end{equation}
or in a matrix form
\begin{equation}
({\mathbf{E}}',\mathbf{B}')=({\mathbf{E}},\mathbf{B})
	\begin{Vmatrix}
a  & b
\\
m  & n                                                    
	\end{Vmatrix}
=
(a{\mathbf{E}}+m\mathbf{B},b{\mathbf{E}}+n\mathbf{B}),                         
\end{equation}
form again a vacuum solution?
Substituting $\mathbf{E}'$ and $\mathbf{B'}$ into Maxwell's vacuum
equations we see that the answer to this question is affirmative iff $m=-b,
n=a$, i.e. iff the corresponding matrix $S$ is of the form
\begin{equation}
S=
	\begin{Vmatrix}
a       & b
\\
-b      & a                                   
	\end{Vmatrix}
.
\end{equation}
The new solution
will have now energy density ${\bf w}'$ and momentum density $\mathbf{S}'$ as
follows:
\[
{\bf w}'=\frac {1}{8\pi}\biggl({\mathbf{E}'}^2 +
			{\mathbf{B}'}^2\biggr)=
\frac {1}{8\pi}(a^2 + b^2)\biggl({\mathbf{E}}^2 + \mathbf{B}^2\biggr),
\]
\[\mathbf{S}'=(a^2+b^2)\frac {c}{4\pi} {\mathbf{E}}\times
			\mathbf{B}.
\]
Obviously, the new and the old solutions will have the same energy and
momentum if $a^2+b^2 =1$, i.e. if the matrix $S$ is {\it unimodular}.
In this case we may put
$a=\cos \alpha$ and $b=\sin \alpha$, where $\alpha=const$,
so transformation (8) becomes
\begin{equation}
\tilde{\mathbf{E}}={\mathbf{E}}\cos \alpha -
				\mathbf{B}\sin \alpha,\        
\tilde{\mathbf{B}}={\mathbf{E}}\sin \alpha + \mathbf{B}\cos
\alpha.
\end{equation}

Transformation (9) is known as {\it electromagnetic duality transformation},
or D-transformation. It has been a subject of many detailed studies in
various aspects and contexts [5]-[8].  It also has greatly
influenced some modern developments in non-Abelian Gauge theories, as well as
some recent general views on duality in field theory, esp. in superstring and
brane theories (classical and quantum).

From physical point of view a basic feature of the D-transformation (9) is,
that the difference between the electric and magnetic fields becomes
non-essential: we may superpose the electric and the magnetic vectors, i.e.
vector-components, of a general electromagnetic field to obtain new
solutions.  From mathematical point of view we see that Maxwell's equations
in vacuum, besides the usual linearity (3) mentioned above, admit also
"cross"-linearity, i.e.  linear combinations of ${\mathbf{E}}$ and $\mathbf{B}$ of a
definite kind determine new solutions.

Any linear map $\phi: \mathbb{R}^2\rightarrow\mathbb{R}^2$, having in the
canonical basis of $\mathbb{R}^2$ a matrix $S$ of the kind (8), is a symmetry
of the canonical complex structure $\mathcal{I}$ of $\mathbb{R}^2$;
we recall that if the canonical basis of $\mathbb{R}^2$ is
denoted by $(\varepsilon^1,\varepsilon^2)$ then ${\cal I}$ is defined by
${\cal I}(\varepsilon^1)=\varepsilon^2$, ${\cal
I}(\varepsilon^2)=-\varepsilon^1$, so if $S$ is given by (8) we have:
$S.{\cal I}.S^{-1}={\cal I}$.  Hence, {\bf the electromagnetic
D-transformations} (9) {\bf coincide with the unimodular symmetries of the
canonical complex structure} ${\cal I}$ {\bf of} $\mathbb{R}^2$.  This
important in our view remark clearly points out that the canonical complex
structure ${\cal I}$ in $\mathbb{R}^2$ should be an {\bf essential element}
of classical electromagnetic theory, so we should in no way neglect it.
Moreover, in my opinion and recalling the above mentioned explicit
introducing of the symplectic structure in hamiltonian mechanics, we {\bf
must find an appropriate way to introduce} ${\cal I}$ {\bf explicitly} in the
equations.

\vskip 0.5cm
\noindent
{\bf Remark 1}. The case with nonzero electric amd magnetic sources will not
be considered here (see lanl e-print: hep-th/0006208).
\vskip 0.5cm

Finally we note that D-transformations change the two well known
invariants: $I_1=(\mathbf{B}^2-{\mathbf{E}}^2)$ and $I_2=2{\mathbf{E}}.\mathbf{B}$ in the
following way:
\begin{align}
& \tilde{I_1}=\tilde{\mathbf{B}}^2-\tilde{\mathbf{E}}^2=
(\mathbf{B}^2-{\mathbf{E}}^2)\cos 2\alpha+
			2{\mathbf{E}}.\mathbf{B}\sin 2\alpha=
I_1\cos 2\alpha+I_2\sin 2\alpha, \\               
&\tilde{I_2}= 2\tilde{\mathbf{E}}.\tilde{\mathbf{B}}=
({\mathbf{E}}^2-\mathbf{B}^2)\sin 2\alpha+
			2{\mathbf{E}}.\mathbf{B}\cos 2\alpha=
-I_1\sin 2\alpha+I_2\cos 2\alpha.                  
\end{align}
It is seen that even the SD-transformation, where $\alpha=\pi/2$, changes
these two invariants:\linebreak $I_1\rightarrow -I_1, \ I_2\rightarrow -I_2$.
This shows that if these two invariants define which solutions should be
called {\it different}, then by making an arbitrary dual transformation we
will always produce different solutions, no matter if these solutions carry
the same energy-momentum or not.
In general we always have
\[
\tilde{I_1}^2+\tilde{I_2}^2=I_1^2+I_2^2,
\]
i.e. the sum of the squared invariants is a D-invariant.

The suggestion coming from the above notices is that the electromagnetic
field, considered as  {\it one physical object}, has {\it two physically
distinguishable interrelated vector components}, $({\mathbf{E}},\mathbf{B})$,
so the adequate mathematical model-object must have two vector components and
must admit 2-dimensional linear transformations of its components, in
particular, the 2-dimensional rotations should be closely related to the
invariance properties of the energy-momentum characteristics of the field.
But every 2-dimensional linear transformation requires a "room where to act",
i.e. a 2-dimensional real vector space has to be {\it explicitly pointed
out} and properly incorporated in theory. This 2-dimensional space has always
been implicitly present inside the electromagnetic field theory, but has
not been given a corresponding respect. Following our earlier papers we
introduce it as follows:
\vskip 0.5cm
	\begin{center}
\hfill\fbox{
	\begin{minipage}{0.65\textwidth}
{\it The electromagnetic field is mathematically represented on
$\mathbb{R}^3$
by an $\mathbb{R}^2$-valued differential 1-form
$\omega$, such that in the canonical basis $(\varepsilon^1,\varepsilon^2)$ in
$\mathbb{R}^2$ the 1-form $\omega$ looks as follows}
\addtocounter{equation}{1}
\begin{equation} \notag
\omega={\mathbf{E}}\otimes \varepsilon^1 +
\mathbf{B}\otimes \varepsilon^2.                
\end{equation}
	\end{minipage}
}
\hfill(\theequation)\\[0.5cm]
	\end{center}
\vskip 0.5cm
\noindent
{\bf Remark 2}. In (12), as well as later on, we identify the vector fields
and 1-forms on $\mathbb{R}^3$ through the euclidean metric and we write, e.g.
$*({\mathbf{E}}\wedge\mathbf{B})=
{\mathbf{E}}\times\mathbf{B}$.
Also, we identify
$(\mathbb{R}^2)^*$ with $\mathbb{R}^2$ through the euclidean metric.
\vskip 0.5cm
Now we have to present equations (1)-(2) correspondingly, i.e. in terms of
$\mathbb{R}^2$-valued objects.

The above assumption (12) requires a general covariance with respect to
transformations in $\mathbb{R}^2$, so, the complex structure $\mathcal{I}$
has to be introduced explicitly in the equations.  In order to do this we
recall that the linear map $\mathcal{I}:\mathbb{R}^2\rightarrow\mathbb{R}^2$
induces a map
\[
{\mathcal{I}_*:\omega\rightarrow \mathcal{I}_*(\omega)=
{\mathbf{E}}\otimes \mathcal{I}(\varepsilon^1)+ \mathbf{B}\otimes
\mathcal{I}(\varepsilon^2)= -\mathbf{B}\otimes \varepsilon^1+
\mathbf{E}}\otimes \varepsilon^2.
\]
We recall also that every operator
$\mathcal{D}$ in the set of differential forms is naturally extended to
vector-valued differential forms according to the rule
$\mathcal{D}\rightarrow \mathcal{D}\times id$,
and $id$ is usually omitted.  Having in mind the
identification of vector fields and 1-forms through the euclidean metric we
introduce now $\mathcal{I}$ in Maxwell's equations (1)-(2) through $\omega$
in the following way:
\begin{equation}
*\mathbf{d}\omega-\frac {1}{c} \frac {\partial }{\partial t}{\cal I}_*
(\omega)=0,\quad \delta\omega=0.                          
\end{equation}
Two other equivalent forms of (13) are given as follows:
\[
{\bf d}\omega-*\frac {1}{c} \frac
{\partial }{\partial t}{\cal I}_* (\omega)=0,\quad \delta \omega =0,
\]
\[
*{\bf d}{\cal I}_*(\omega)+
\frac {1}{c} \frac {\partial }{\partial t}\omega=0, \quad \delta \omega =0.
\]
\noindent
In order to verify the equivalence of (13) to Maxwell equations (1)-(2)
we compute the marked operations.  We obtain
\[
*{\bf d}\omega-
\frac {1}{c} \frac {\partial }{\partial t}{\cal I}_*(\omega)=
\left({\rm curl}\,\mathbf{E}+\frac1c\frac{\partial \mathbf{B}}{\partial
t}\right) \otimes\varepsilon^1+ \left({\rm curl}\,\mathbf{B}-
\frac1c\frac{\partial {\mathbf{E}}}{\partial t}\right)\otimes\varepsilon^2,
\]
The second equation $\delta \omega=0$ is, obviously,
equivalent to
$$
{\rm div}\,{\mathbf{E}}\otimes \varepsilon^1 +
{\rm div}\,\mathbf{B}\otimes \varepsilon^2=0
$$
since $\delta=-{\rm div}$. Hence, (13) coincides with (1)-(2).

We shall emphasize once again that according to our general assumption (12)
the field $\omega$ will have different representations in the different bases
of $\mathbb{R}^2$.
Changing the basis $(\varepsilon^1,\varepsilon^2)$ to any other basis
$\varepsilon^{1'}=\varphi(\varepsilon^1),
\varepsilon^{2'}=\varphi(\varepsilon^2)$,
means, of course, that in equations (13) the field $\omega$ changes to
$\varphi_*\omega$ and the complex structure ${\cal I}$
changes to $\varphi{\cal I}\varphi^{-1}$. In some sense this means that we
have two fields now: $\omega$ and ${\cal I}$, but ${\cal I}$ is given
beforehand and it is not determined by equations (13). So, in the new basis
the $\mathcal{I}$-dependent equations of (13) will look like
\[
*{\bf d}\varphi_*\omega-
\frac1c\frac{\partial }{\partial t}
(\varphi{\cal I}\varphi^{-1})_*(\varphi_*\omega)=0.
\]
If $\varphi$ is a symmetry of ${\cal I}:
\varphi{\cal I}\varphi^{-1}={\cal I}$, then we transform just $\omega$ to
$\varphi_*\omega$.

In order to write down the Poynting energy-momentum balance relation we
recall the product of vector-valued differential forms. Let
$\Phi= \Phi^a\otimes e_a$ and $\Psi= \Psi^b\otimes k_b$ are two differential
forms on some manifold with values in the vector spaces $V_1$ and $V_2$
with bases $\{e_a\}, a=1,...,n$ and $\{k_b\}, b=1,...,m$, respectively.
Let $f:V_1\times V_2\rightarrow W$ is a bilinear map valued in a third vector
space $W$.  Then a new differential form, denoted by $f(\Phi,\Psi)$, on the
same manifold and valued in $W$ is defined by
\[
f(\Phi,\Psi)=\Phi^a\wedge \Psi^b\otimes f(e_a,k_b).
\]
Clearly, if the original forms are $p$ and $q$ respectively,
then the product is a $(p+q)$-form.

Assume now that $V_1=V_2=\mathbb{R}^2$ and the bilinear map is the
exterior product:\linebreak
$\wedge:\mathbb{R}^2\times \mathbb{R}^2\rightarrow
\Lambda^2(\mathbb{R}^2)$.

Let's compute the expression $\wedge(\omega,{\bf d}\omega)$.
\begin{align*}
& \wedge(\omega,{\bf d}\omega)=\wedge({\mathbf{E}}\otimes \varepsilon^1+
\mathbf{B}\otimes \varepsilon^2,{\bf d}{\mathbf{E}}\otimes \varepsilon^1+
{\bf d}\mathbf{B}\otimes \varepsilon^2)=
({\mathbf{E}}\wedge{\bf d}\mathbf{B}-
\mathbf{B}\wedge{\bf d}{\mathbf{E}})\otimes
\varepsilon^1\wedge\varepsilon^2
\\
& =-{\bf
d}({\mathbf{E}}\wedge\mathbf{B})\otimes\varepsilon^1\wedge\varepsilon^2=
-{\bf d}(**({\mathbf{E}}\wedge\mathbf{B}))\otimes\varepsilon^1\wedge\varepsilon^2=
*\delta({\mathbf{E}}\times\mathbf{B})\otimes\varepsilon^1\wedge\varepsilon^2
\\
& =-*{\rm
div}({\mathbf{E}}\times\mathbf{B})\otimes\varepsilon^1\wedge\varepsilon^2=
-{\rm div}({\mathbf{E}}\times\mathbf{B})dx\wedge dy\wedge dz\otimes
\varepsilon^1\wedge\varepsilon^2.
\end{align*}
Following the same rules we obtain
\[
\wedge\left(\omega,*\frac1c\frac{\partial }{\partial t}
{\cal I}_*\omega\right)=
\frac1c\frac{\partial }{\partial t} \frac{{\mathbf{E}}^2+
\mathbf{B}^2}{2}dx\wedge dy\wedge
dz\otimes\varepsilon^1\wedge\varepsilon^2,
\]
So, the Poynting energy-momentum
balance relation is given by
\begin{equation}
\wedge\left(\omega, {\bf d}\omega-*\frac1c\frac{\partial }{\partial t}
{\cal I}_*\omega\right)=0.                                    
\end{equation}
Since the orthonormal 2-form $\varepsilon^1\wedge\varepsilon^2$ is invariant
with respect to rotations (and even with respect to unimodular
transformations in $\mathbb{R}^2$) we have the duality invariance of the
above energy-momentum quantities and relations.

Note the following simple forms of the energy density
\[
\frac{1}{8\pi}*\wedge\left(\omega,*{\cal I}_*\omega\right)=
\frac{{\mathbf{E}}^2+\mathbf{B}^2}{8\pi}\varepsilon^1\wedge\varepsilon^2,
\]
and of the Poynting vector,
\[
\frac{c}{8\pi}*\wedge(\omega,\omega)=
\frac{c}{4\pi}{\mathbf{E}}\times\mathbf{B}
\otimes\varepsilon^1\wedge\varepsilon^2,
\]
the D-invariance is obvious. As for the general $\mathbb{R}^2$ covariance of
the second equation of (13) it is obvious.

Resuming, we may say that pursuing the mathematical adequacy
of the correspondence:  {\it one physical object - one mathematical
model-object}, we came to the idea to introduce the $\mathbb{R}^2$-valued
1-form $\omega$ as the mathematical model-field.  This, in turn, set the
problem for general $\mathbb{R}^2$ covariance of the equations and this
problem was solved through introducing explicitly the canonical complex
structure $\mathcal{I}$ in the dynamical equations (13) of the theory. This
means that an arbitrary linear transformation
\[
{\mathbf{E}}'=a{\mathbf{E}}+m\mathbf{B},\
\mathbf{B}'=b{\mathbf{E}}+n\mathbf{B},
\]
will give again a solution of (13).

\vskip 0.5cm

\section {4-Dimensional Consideration}
\subsection{Classical Electrodynamics}
We are going to reveal the complex structure in the 4-dimensional
formulation of Maxwell equations in two ways. The first way is quite direct
and consists in the following.

In the 4-dimensional formulation of Maxwell equations we consider the time
variable $x^4=\xi=ct$, where $c$ is the velocity of light, as a coordinate and
treat it in the same way as the other three spatial coordinates
$(x^1,x^2,x^3=x,y,z)$.
So, the base manifold becomes 4-dimensional, in fact, $\mathbb{R}^4$. Since
the vector fields on $\mathbb{R}^4$ form a 4-dimensional module,
$\mathbf{E}$ and $\mathbf{B}$ can not be considered in general as vector
fields on $\mathbb{R}^4$. But the couple $(\mathbf{E},\mathbf{B})$ has 6
components, therefore we consider the space $\Lambda^2(\mathbb{R}^4)$ of
2-forms, which is a 6-dimensional module, as a natural solution space.
Moreover, as it is well known, in the basis
$$
 dx\wedge dy,  \ \ dx\wedge dz,  \ \ dy\wedge dz,  \ \
 dx\wedge d\xi,  \ \ dy\wedge d\xi,  \ \ dz\wedge d\xi
$$
of $\Lambda^2(\mathbb{R}^4)$ if we put for $F\in\Lambda^2(\mathbb{R}^4)$
\[
F_{12}=\mathbf{B}^3, \ \ F_{13}=-\mathbf{B}^2, \ \ F_{23}=\mathbf{B}^1, \ \
F_{14}=\mathbf{E}^1, \ \ F_{24}=\mathbf{E}^2, \ \ F_{34}=\mathbf{E}^3,
\]
then the equation $\mathbf{d}F=0$ gives the first couple (1) of Maxwell
equations. Similarly, if we consider the 2-form $\Phi(F)$, given in this
basis by
\begin{alignat*}{3}
& (\Phi F)_{12}=\mathbf{E}^3, &\quad (\Phi F)_{13}=-\mathbf{E}^2,
&\quad (\Phi F)_{23}=\mathbf{E}^1; \\
& (\Phi F)_{14}=-\mathbf{B}^1,&\quad (\Phi F)_{24}=-\mathbf{B}^2,
&\quad (\Phi F)_{34}=-\mathbf{B}^3,
\end{alignat*}
then the equation $\mathbf{d}(\Phi F)=0$ gives the second couple (2)
of Maxwell equations. Hence, Maxwell equations in vacuum become
\begin{equation}
\mathbf{d}F=0,\ \mathbf{d}(\Phi F)=0.                      
\end{equation}
We especially note that {\it no pseudometric is needed} to write down
these equations. We also note that any two linear combinations
\begin{equation}
F'=aF+b(\Phi F),\ \ (\Phi F)'=mF+n(\Phi F)                  
\end{equation}
with arbitrary $(a,b,m,n)$ define again a solution
$\left(F',(\Phi F)'\right)$.

The linear map
$\Phi:\Lambda^2(\mathbb{R}^4)\rightarrow \Lambda^2(\mathbb{R}^4)$,
as defined above, has the property
$$
\Phi_\circ \Phi=-id_{\Lambda^2(\mathbb{R}^4)},
$$
hence {\it it introduces complex structure} in the space
$\Lambda^2(\mathbb{R}^4)$.

The second way follows the considerations in the 3-dimensional case and makes
use of the complex structure $\mathcal{I}$ of $\mathbb{R}^2$, and of the
space $\Lambda^2(\mathbb{R}^4,\mathbb{R}^2)$ of $\mathbb{R}^2$-valued
differential 2-forms on $\mathbb{R}^4$. In general an $\mathbb{R}^2$ valued
2-form $\Omega$ on $\mathbb{R}^4$ looks as follows:
\[
\Omega=F_1\otimes\varepsilon^1+F_2\otimes \varepsilon^2.
\]
Consider now the two linear maps:
\[
\mathcal{F}:\Lambda^2(\mathbb{R}^4)\rightarrow\Lambda^2(\mathbb{R}^4),
\ \quad\varphi:\mathbb{R}^2\rightarrow\mathbb{R}^2.
\]
These maps induce a map
$(\mathcal{F},\varphi):
\Lambda^2(\mathbb{R}^4,\mathbb{R}^2)
\rightarrow\Lambda^2(\mathbb{R}^4,\mathbb{R}^2)$ by the rule:
\[
(\mathcal{F},\varphi)(\Omega)=
(\mathcal{F},\varphi)({\bf F}_a\otimes\varepsilon^a)=
\mathcal{F}({\bf F}_a)\otimes\varphi(\varepsilon^a),\quad
\text{summation over $a$=1,2}.
\]
It is natural to ask now is it possible the joint action of these two maps to
keep $\Omega$ unchanged, i.e. to have
$$
(\mathcal{F},\varphi)(\Omega)=\Omega.
$$
In such a case the form $\Omega$ is called
$(\mathcal{F},\varphi)$-{\it equivariant}.  If $\varphi$ is a linear
isomorphism and we identify $\mathcal{F}$ with
$(\mathcal{F},id_{\mathbb{R}^2})$ and $\varphi$ with
$(id_{\Lambda^2(\mathbb{R}^4)},\varphi)$, we can equivalently write
\[
\mathcal{F}(\Omega)=\varphi^{-1}(\Omega).
\]
If we specialize now: $\varphi=\mathcal{I}$ we readily find that
the $(\mathcal{F},\mathcal{I})$-equivariant forms $\Omega$ must satisfy
\[
(\mathcal{F},\mathcal{I})(\Omega)=
-\mathcal{F}(F_2)\otimes\varepsilon^1+\mathcal{F}(
F_1)\otimes\varepsilon^2= F_1\otimes\varepsilon^1+
F_2\otimes\varepsilon^2=\Omega.
\]
Hence, we must have $\mathcal{F}(F_1)=F_2$ and $\mathcal{F}
(F_2)=-F_1$, i.e. $\mathcal{F}_\circ\mathcal{F}=-id$.
In other words, the property
$\mathcal{I}_\circ\mathcal{I}=-id_{\mathbb{R}^2}$ is carried
over to $\mathcal{F}$:
$\mathcal{F}_\circ\mathcal{F}=-id_{\Lambda^2(\mathbb{R}^4)}$.

Hence, recalling the linear map $\Phi$, introduced above, and
working with $(\Phi,\mathcal{I})$-equivariant
2-forms on $\mathbb{R}^4$, we can replace the action of $\mathcal{I}$ with
the action of $\Phi$.  And that's why in the 4-dimensional formulation of
electrodynamics we have general $\mathbb{R}^2$ covariance in the sense of
(16) if we work with forms $\Omega$ of the kind
$\Omega=F\otimes\varepsilon^1+\Phi F\otimes\varepsilon^2$.
In the 3-dimensional formulation this is not possible to be done since we
work there on $\mathbb{R}^3$ and no map
$\Phi:\Lambda^1(\mathbb{R}^3)\rightarrow\Lambda^1(\mathbb{R}^3)$ with the
property $\Phi_\circ\Phi=-id_{\Lambda^1(\mathbb{R}^3)}$ exists since
$\Lambda^1(\mathbb{R}^3)$ is a 3-dimensional space and the
relation $\Phi_\circ\Phi=-id$ requires {\it even}-dimensional space, so we
have to introduce the complex structure through $\mathbb{R}^2$ only.

Having in view these considerations our basic assumption for the
algebraic nature of the mathematical-model object must read:
\vskip 0.5cm
	\begin{center}
\hfill\fbox{
	\begin{minipage}{0.65\textwidth}
{\it The electromagnetic field is mathematically represented on}
$\mathbb{R}^4$ {\it by a}
$(\Phi,\mathcal{I})$-{\it equivariant} $\mathbb{R}^2$ {\it valued
2-form} $\Omega$ {\it such that in the canonical basis
$(\varepsilon^1,\varepsilon^2)$ in
$\mathbb{R}^2$ the 1-form $\Omega$ looks as follows}
\addtocounter{equation}{1}
\begin{equation} \notag
\Omega= F\otimes\varepsilon^1+ (\Phi F)\otimes\varepsilon^2.       
\end{equation}
	\end{minipage}
}
\hfill(\theequation)\\[0.5cm]
	\end{center}
The pure field Maxwell equations, expressed through the
$(\Phi,\mathcal{I})$-equivariant 2-form $\Omega$
have, obviously, general $\mathbb{R}^2$ covariance and are equivalent to
\begin{equation}
{\bf d}\Omega=0.                               
\end{equation}

Now we are going to show how the well known from standard relativistic
electrodynamics on Minkowski space-time pseudometric structures can be
introduced by means of the complex structure $\Phi$, which is the {\it only
external mathematical structure} on $\mathbb{R}^4$ introduced by Maxwell
equations.

We begin with giving the matrix of $\Phi$ in the above given coordinate basis
$dx^\mu\wedge dx^\nu, \ \mu<\nu$, where $x^4=\xi=ct$,
of the space $\Lambda^2(\mathbb{R}^4)$ considered as a 6-dimensional module
over the algebra  $C^{\infty}(\mathbb{R}^4)$ of all smooth real valued
functions on $\mathbb{R}^4$.
\begin{equation}
\Phi^{(\alpha\beta)}_{(\mu\nu)}=\begin{Vmatrix}
0  & 0  & 0  & 0  & 0  & -1\\
0  & 0  & 0  & 0  & 1  & 0 \\
0  & 0  & 0  & -1 & 0  & 0 \\
0  & 0  & 1  & 0  & 0  & 0 \\                            
0  & -1 & 0  & 0  & 0  & 0 \\
1  & 0  & 0  & 0  & 0  & 0 \\
\end{Vmatrix},
\end{equation}
where $\alpha<\beta$, $\mu<\nu$, $(\alpha\beta)$ numbers the rows and
$(\mu\nu)$ numbers the columns. Hence,
\begin{alignat*}{3}
\Phi(dx\wedge dy) & =-dz\wedge d\xi, & \quad
\Phi(dx\wedge dz) & =dy\wedge d\xi, & \quad
\Phi(dy\wedge dz) & =-dx\wedge d\xi, \\
\Phi(dx\wedge d\xi) & =dy\wedge dz, & \quad
\Phi(dy\wedge d\xi) & =-dx\wedge dz, & \quad
\Phi(dz\wedge d\xi) & =dx\wedge dy.
\end{alignat*}
\vskip 0.3cm
\noindent
{\bf Remark 3}.
If we introduce the notation $dx\wedge dy =e^1,\dots,\ dz\wedge d\xi=e^6$,
and $\frac{\partial}{\partial x}\wedge \frac{\partial}{\partial y} =e_1,
\dots,\frac{\partial}{\partial z}\wedge \frac{\partial }{\partial \xi}
=e_6$, so that $\{e^i\}$ and $\{e_j\}$ are
dual bases, then $\Phi$ may be considered as $(2,2)$ tensor and represented
by
\[
\Phi=\sum _{\substack{1\le i\le 6}}(-1)^i e_i\otimes e^{7-i}.
\]

In order to build all additional structures needed and used in
Electrodynamics we are going to make use of the {\it Poincar\'e}
isomorphism $\mathfrak{P}$, and for convenience we recall
its algebraical constuction [10].
Let $\Lambda^p(V)$ be the space of $p$-vectors, i.e. fully antisymmetric
contravariant $p$-tensors, and
$\Lambda^{(n-p)}(V^*)$ be the space of $(n-p)$-forms over the pair of dual
$n$-dimensional linear spaces $(V,V^*)$,
and $p=1,\dots,n$.  If $\{e_i\}$ and $\{\varepsilon^{i}\}$ are two dual bases
we have the $n$-vector $\omega=e_1\wedge \dots \wedge e_n$ and the
$n$-form $\omega^*=\varepsilon^1\wedge \dots \wedge \varepsilon^n$. The
duality requires $\langle\varepsilon^i,e_j\rangle=\delta^i_j$ and
$\langle\omega^*,\omega\rangle=1$.  If $x\in V$ and $\alpha\in
\Lambda^p(V^*)$ then by means of the insertion operator $i(x)$ we obtain the
$(p-1)$-form $i(x)\alpha$: if $\alpha$ is decomposable:
$\alpha=\alpha^1\wedge \alpha^2\wedge \dots \wedge\alpha^p$, where
$\alpha^1,\dots,\alpha^p$ are 1-forms, then
\begin{align*}
\begin{split}
i(x)\alpha=&\langle\alpha^1,x\rangle\alpha^2\wedge \dots \wedge \alpha^p
-\langle\alpha^2,x\rangle\alpha^1\wedge \alpha^3\wedge \dots \wedge \alpha^p
+\dots  \\
&+(-1)^{p-1}\langle\alpha^p,x\rangle\alpha^1\wedge \dots\wedge \alpha^{p-1}.
\end{split}
\end{align*}
Now let $x_1\wedge x_2\wedge \dots \wedge x_p \in \Lambda^p(V)$. In the
decomposable case the Poincar\'e isomorphism $\mathfrak{P}$ acts as follows:
\begin{equation}
\mathfrak{P}_p(x_1\wedge x_2\wedge\dots\wedge x_p)=
i(x_p)_\circ i(x_{p-1})_\circ \dots _\circ i(x_1)\omega^*.      
\end{equation}
In the same way
\begin{equation}
\mathfrak{P}^p(\alpha^1\wedge \dots \wedge\alpha^p)=               
i(\alpha^p)_\circ \dots _\circ i(\alpha^1)\omega,
\end{equation}
where
\[
\begin{split}
i(\alpha^i)(x^1\wedge x^2\wedge \dots x^p)=
&\langle\alpha^i,x^1\rangle x^2\wedge \dots \wedge x^p
-\langle\alpha^i,x^2\rangle x^1\wedge x^3\wedge \dots \wedge x^p
+\dots  \\
&+(-1)^{p-1}\langle\alpha^i,x^p\rangle x^1\wedge \dots\wedge x^{p-1}.
\end{split}
\]
For nondecomposable p-vectors $\mathfrak{P}$ is extended by linearity. We note
also the relations:
\[
\mathfrak{P}_{n-p}\mathfrak{P}^p=(-1)^{p(n-p)}id,\quad
\mathfrak{P}^{n-p}\mathfrak{P}_p=(-1)^{p(n-p)}id,
\]
\[
\langle\mathfrak{P}^p(\alpha^1\wedge \dots \wedge \alpha^p),
\mathfrak{P}_p(x_1\wedge \dots \wedge x_p)\rangle=
\langle\alpha^1\wedge \dots \wedge \alpha^p,x_1\wedge \dots \wedge x_p\rangle.
\]
On an arbitrary basis element $\mathfrak{P}$ acts in the following way:
\begin{equation}
\mathfrak{P}^p(\varepsilon^{i_1}\wedge \dots \wedge \varepsilon^{i_p})=
(-1)^{\mbox{\small$\displaystyle\sum_{k=1}^p (i_k -k)$}}e_{i_{p-1}}   
\wedge \dots \wedge e_{i_n},
\end{equation}
where $i_1<\dots<i_p$ and $i_{p+1}<\dots<i_n$ are complementary
$p-$ and $(n-p)-$ tuples.

On the manifold $\mathbb{R}^4$ we have the dual bases $\{dx^i\}$ and
$\left\{\frac{\partial}{\partial x^i}\right\}$ of the two modules of 1-forms
and vector fields. The corresponding $\omega$ and $\omega^*$ are
\[
\omega=\frac{\partial}{\partial x}\wedge\frac{\partial}{\partial y}\wedge
\frac{\partial}{\partial z}\wedge\frac{\partial}{\partial \xi},\quad
\omega^*=dx\wedge dy\wedge dz\wedge d\xi.
\]

Now we define the nondegenerate operator $\mathfrak{D}:
\Lambda^2(T^*\mathbb{R}^4)\rightarrow\Lambda^2(T\mathbb{R}^4)$ as follows:
\begin{equation}
\mathfrak{D}=-\mathfrak{P}_\circ\Phi,                          
\end{equation}
where $\Phi$ is defined in (19). On the basis elements $\mathfrak{D}$ acts in
the following way:
\[
\mathfrak{D}(dx\wedge dy)=\frac{\partial}{\partial x}\wedge
\frac{\partial}{\partial y}, \quad
\mathfrak{D}(dx\wedge dz)=\frac{\partial}{\partial x}\wedge
\frac{\partial}{\partial z}, \quad
\mathfrak{D}(dy\wedge dz)=\frac{\partial}{\partial y}\wedge
\frac{\partial}{\partial z}
\]
\[
\mathfrak{D}(dx\wedge d\xi)=-\frac{\partial}{\partial x}\wedge
\frac{\partial}{\partial \xi}, \quad
\mathfrak{D}(dy\wedge d\xi)=-\frac{\partial}{\partial y}\wedge
\frac{\partial}{\partial \xi}, \quad
\mathfrak{D}(dz\wedge d\xi)=-\frac{\partial}{\partial z}\wedge
\frac{\partial}{\partial \xi}.
\]
Note that
\[
\mathfrak{D}(dx\wedge dy)\wedge\mathfrak{D}(dz\wedge d\xi)=-\omega.
\]
The isomorphism $\mathfrak{D}$ defines a bilinear form
$h^2:\Lambda^2(T^* \mathbb{R}^4)\times \Lambda^2(T^*\mathbb{R}^4)
\rightarrow \mathbb{R}$
according to the rule:
$h^2(\alpha,\beta)=\langle\mathfrak{D}(\alpha),\beta\rangle$.  In our
coordinate basis $h^2$ has components as follows:
$$
h^{2(12,12)}=h^{2(13,13)}=h^{2(23,23)}
=-h^{2(14,14)}=-h^{2(24,24)}=-h^{2(34,34)}=1,
$$
and all other components are equal to zero.
On the other hand the complex structure $\Phi$ defines a bilinear form
$\tilde{h^2} :\Lambda^2(T^* \mathbb{R}^4)\times\Lambda^2(T^*\mathbb{R}^4)
\rightarrow \mathbb{R}$ according to the rule
$$
\alpha\wedge (\Phi \beta)=-\tilde{h^2}(\alpha,\beta)\omega^*.
$$
Making use of the canonical basis $dx^\mu\wedge dx^\nu,\ \mu<\nu$, it is easy
to show that {\bf these two bilinear forms coincide}:  $h^2=\tilde h^2$.
For example, for $\tilde h^{2(12,12)}$ we obtain
\[
(dx\wedge dy)\wedge \Phi(dx\wedge dy)=-dx\wedge dy\wedge dz\wedge d\xi=
-\omega^*=-\tilde h^2(dx\wedge dy,dx\wedge dy)\omega^*,
\]
hence, $\tilde h^{2(12,12)}=1=h^{2(12,12)}$.

We are going to extend the complex structure $\Phi$ to a map
 $\mathbf{\circledast}$:  $\Lambda^p(T^*
\mathbb{R}^4)\rightarrow\Lambda^{4-p}(T^*\mathbb{R}^4), \linebreak
p=1,\dots,4$. To this end we first prove the following

{\bf Proposition 1.}
There exists unique (up to a
sign) linear isomorphism
$\varphi: T_x^*(\mathbb{R}^4)\rightarrow T_x(\mathbb{R}^4),
\ x\in \mathbb{R}^4$, satisfying the two conditions:

	1. $\varphi$ is represented by a diagonal matrix in the bases
$\{dx^i\}$, $\{\frac{\partial}{\partial x^i}\}$,

	2. $\wedge^2\varphi=\mathfrak{D}$.

{\bf Proof}. Actually, these two conditions imply
$$
\varphi(dx^\mu)=\lambda^\mu\frac{\partial}{\partial x^\mu},\ \  \text{no
summation over $\mu$},
$$
$$
(\wedge^2\varphi)^{\mu\nu,\alpha\beta}=
\varphi^{\mu\alpha}\varphi^{\nu\beta}-\varphi^{\mu\beta}\varphi^{\nu\alpha}=
\mathfrak{D}^{\mu\nu,\alpha\beta},
\quad \mu<\nu,\ \alpha<\beta.
$$
The diagonal form of $\varphi^{\mu\nu}$ reduces the components of
$\wedge^2\varphi$
to $ \varphi^{\mu\mu}\varphi^{\nu\nu}=\lambda^\mu \lambda^\nu$,
$\mu<\nu$.
Now, in view of the components of $\mathfrak{D}$ we obtain:
$$
-\lambda^1=-\lambda^2=-\lambda^3=\lambda^4=1,\  \text{or}\quad
\lambda^1=\lambda^2=\lambda^3=-\lambda^4=1.
$$
\noindent
This ends the proof. We are going to work further with the first of these two
solutions.

Thus we have $\varphi$ and
$\wedge^2\varphi$. Computing $\Lambda^3\varphi$ in the bases
$$
dx^\alpha\wedge dx^\beta\wedge dx^\mu,\quad
\frac{\partial}{\partial x^\alpha}
\wedge\frac{\partial}{\partial x^\beta}
\wedge\frac{\partial}{\partial x^\mu},
\ \ \alpha<\beta<\mu,
$$
we find that $\wedge^3\varphi$ has only diagonal components equal to
$(-1,1,1,1)$.  Finally, $\wedge^4\varphi$ has only one component
$(\wedge^4\varphi)^{1234}=-1$, so that $\wedge^4\varphi(\omega^*)=-\omega$.

In the well known way the linear isomorphisms $\wedge^p\varphi,\
p=1,\dots,4$, define bilinear forms $h^p$ in $\Lambda^p(T^*\mathbb{R}^4)$ by
the rule $h^p(\alpha,\beta)=\langle\wedge^p\varphi(\alpha),\beta\rangle$, and
all these four bilinear forms are {\it nondegenerate}.

Now we extend $\Phi$ to $\mathbf{\circledast}$ by the rule
\begin{equation}
\alpha^p\wedge \mathbf{\circledast}_p\beta^p=                
-h^p(\alpha^p,\beta^p)\omega^*, \ p=1,2,3,4 ,
\end{equation}
where $\alpha^p$ and $\beta^p$ are p-forms, and $\circledast_p$ means the
restriction of $\circledast$ to p-forms. The above algebraic considerations
show that:

1. The bilinear forms $h^p$ define {\bf pseudoeuclidean metric structures} in
the spaces \linebreak
$\Lambda^p(T^*\mathbb{R}^4), \ p=1,\dots,4$, with corresponding signitures:
$$
(-1,-1,-1,1),\ (1,1,1,-1,-1,-1),\ (-1,1,1,1),\ (-1),
$$
and in fact, in all tensor bundles over the manifold $\mathbb{R}^4$.

2. The following relation holds:
\begin{equation}
\mathbf{\circledast}_p=-\mathfrak{P}_\circ\wedge^p\varphi, \ p=1,2,3,4.
\end{equation}
For $p=2$ relation (25) is obvious in view of relation
(23). We illustrate it for the case $p=1$.
\[
-\mathfrak{P}_o\wedge^1\varphi(dx)=
-\mathfrak{P}_o\varphi(dx)
=\mathfrak{P}\left(\frac{\partial}{\partial x}\right)=
dy\wedge dz\wedge d\xi.
\]
So, according to (24) we obtain
\[
dx\wedge \circledast_1{dx}=
-h^1(dx,dx)dx\wedge dy\wedge dz\wedge d\xi=
-(-1)dx\wedge dy\wedge dz\wedge d\xi=
dx\wedge(-\mathfrak{P}_o\wedge^1\varphi(dx)).
\]
Clearly, the operator $\circledast$ coinsides with the Hodge $*$ operator,
defined by the bilinear forms $h^p$, confirming our assertion that we may
come to all known metric structures starting with the complex structure
$\Phi$ and making use of the Poincar\'e isomorphism $\mathfrak{P}$.

\subsection{Symmetries of Maxwell equations}
In order to find the symmetries of Maxwell equations (15), generated by
vector fields $X$ on the base manifold $\mathbb{R}^4$ we have to solve the
equation $L_{X}\Phi=0$, where $L_{X}$ is the Lie derivative along $X$, and
this is due to the fact that the Lie derivative commutes with the exterior
derivative.  We shall solve this problem as a particular case of the more
general problem to find the symmetries of the operator
$\circledast$, i.e.  to find those $X$ along
which we have
$$
\left[L_X,\mathbf{\circledast}\right]\equiv
\left(L_X\right)_\circ \mathbf{\circledast}- \mathbf{\circledast}_\circ
\left(L_X\right)=0.
$$
First we note
\[
L_{X}\left(\alpha\wedge\mathbf{\circledast}\beta\right)=
\left(L_X \alpha\right)\wedge {\circledast}\beta + \alpha\wedge L_X\left(
\mathbf{\circledast}\beta\right)=
\left(L_X \alpha\right)\wedge {\circledast}\beta+
\alpha\wedge\left[L_X,\mathbf{\circledast}\right]\beta
+\alpha\wedge\mathbf{\circledast}L_X\beta.
\]
On the other hand, making use of relation (24), we obtain
$(\alpha,\beta\in \Lambda^p(T^*\mathbb{R}^4))$
\begin{align*}
& L_{X}\left(\alpha\wedge\mathbf{\circledast}\beta\right)=
-(L_Xh^p)(\alpha,\beta)\omega^* - h^p(L_X\alpha,\beta)\omega^*-
h^p(\alpha,L_X\beta)\omega^*-
h^p(\alpha,\beta)L_X\omega^* \\
&=-(L_Xh^p)(\alpha,\beta)\omega^*+
\left(L_X \alpha\right)\wedge\mathbf{\circledast}\beta
+\alpha\wedge\mathbf{\circledast}L_X\beta -h^p(\alpha,\beta){\rm div}X.
\omega^* \\
&=-(L_Xh^p)(\alpha,\beta)\omega^*+
L_{X}\left(\alpha\wedge\mathbf{\circledast}\beta\right)-
\alpha\wedge\left[L_X,\mathbf{\circledast}\right]\beta-
h^p(\alpha,\beta){\rm div}X.\omega^*.
\end{align*}
Since $\alpha$ and $\beta$ are arbitrary $p$-forms from this relation it
follows that $\left[L_X,\mathbf{\circledast}\right]=0$ iff
\begin{equation}
L_Xh^p=-{\rm div}X.h^p,\quad p=1,2,3,4.                
\end{equation}
\noindent
{\bf Remark}. Relation (26) shows that the local symmetries of
$\circledast_p$ are conformal symmetries of special kind of the metric $h^p$,
namely, the conformal multiplyer generated by the vector field $X$, is equal
to $-{\rm div}X$.

For the case we are interesting in, we obtain
\begin{equation}
L_X\Phi=0\Longleftrightarrow L_Xh^2=-{\rm div}X.h^2.       
\end{equation}
From this relation we obtain the following (independent)
equations for the components of any local symmetry $X$ of $\Phi$.

\begin{alignat*}{2}
 2\left(\frac{\partial X^1}{\partial x}+\frac{\partial X^2}{\partial y}\right)
& ={\rm div}X, & \qquad
2\left(\frac{\partial X^1}{\partial x}+\frac{\partial X^1}{\partial
\xi}\right)
& ={\rm div}X,  \qquad \\
 2\left(\frac{\partial X^1}{\partial x}+\frac{\partial X^3}{\partial z}\right)
& ={\rm div}X, & \qquad
 2\left(\frac{\partial X^2}{\partial y}+\frac{\partial X^4}{\partial
\xi}\right)
& ={\rm div}X,  \qquad  \\
 2\left(\frac{\partial X^2}{\partial y}+\frac{\partial X^3}{\partial z}\right)
& ={\rm div}X, & \qquad
 2\left(\frac{\partial X^3}{\partial z}+\frac{\partial X^4}{\partial
\xi}\right)
& ={\rm div}X,  \qquad \\
\left(\frac{\partial X^2}{\partial x}+\frac{\partial X^1}{\partial y}\right)
& =0, & \qquad
\left(\frac{\partial X^4}{\partial x}-\frac{\partial X^1}{\partial \xi}\right)
& =0,  \qquad  \\
\left(\frac{\partial X^3}{\partial x}+\frac{\partial X^1}{\partial z}\right)
& =0, & \qquad
\left(\frac{\partial X^4}{\partial y}-\frac{\partial X^2}{\partial \xi}\right)
& =0,  \qquad \\
\left(\frac{\partial X^3}{\partial y}+\frac{\partial X^2}{\partial z}\right)
& =0, & \qquad
\left(\frac{\partial X^4}{\partial z}-\frac{\partial X^3}{\partial \xi}\right)
&=0.
\end{alignat*}

These equations have the following solutions:

1. Translations:
$$
X=\frac{\partial}{\partial x},\quad X=\frac{\partial}{\partial y},\quad
X=\frac{\partial}{\partial z},\quad X=\frac{\partial}{\partial \xi},
$$
as well as any linear combination with {\it constant} coefficients of
these four vector fields;
\vskip 0.4cm
2. Spatial rotations:
\[
X=y\frac{\partial}{\partial x}-x\frac{\partial}{\partial y},\quad
X=z\frac{\partial}{\partial y}-y\frac{\partial}{\partial z},\quad
X=x\frac{\partial}{\partial z}-z\frac{\partial}{\partial x};
\]
\vskip 0.4cm
3. Space-time rotations:
\[
X=x\frac{\partial}{\partial \xi}+\xi\frac{\partial}{\partial x},\quad
X=y\frac{\partial}{\partial \xi}+\xi\frac{\partial}{\partial y},\quad
X=z\frac{\partial}{\partial \xi}+\xi\frac{\partial}{\partial z}
\]
\vskip 0.4cm
4. Dilatations:
\[
X=x\frac{\partial}{\partial x}+y\frac{\partial}{\partial y}+
z\frac{\partial}{\partial z}+\xi\frac{\partial}{\partial \xi},\quad \text{or}
\quad X=x^\mu\frac{\partial}{\partial x^\mu};
\]
\vskip 0.4cm
5. Special conformal (with respect to the bilinear form $h^2$) vector fields:
\[
X_\mu=\left(h^1_{\alpha\beta} x^\alpha x^\beta\right)
\frac{\partial}{\partial x^\mu}-2h^1_{\mu\nu}x^\nu\left(x^\sigma
\frac{\partial}{\partial x^\sigma}\right),\quad \mu=1,\dots, 4.
\]

Let's consider the flows generated by the above vector
fields.

1. The translation vector fields generate flows as follows:
\[
x^{\mu'}=x^\mu+a^\mu, \ \ a^\mu \ \text{are 4 constants}.
\]
\vskip 0.4cm
2. The spatial rotations generate "rotational" flows inside the three planes
$(x,y)$, $(x,z)$ and $(y,z)$ as follows:
\begin{alignat*}{3}
x'&=x\,\cos(s)+y\,\sin(s)&,\quad x'&=x\,\cos(s)+z\,\sin(s)&,\quad
y'&=y\,\cos(s)+z\,\sin(s) \\
y'&=-x\,\sin(s)+y\,\cos(s)&,\quad
z'&=-x\,\sin(s)+z\,\cos(s)&\quad z'&=-y\,\sin(s)+z\,\cos(s).
\end{alignat*}

\vskip 0.4cm
3. The space-time rotations generate the following flows:
\begin{alignat*}{3}
x'& =x\,{\rm ch}(s)+\xi\,{\rm sh}(s),&\quad y'& =y\,{\rm ch}(s)+\xi\,{\rm
sh}(s),&\quad z'& =z\,{\rm ch}(s)+\xi\,{\rm sh}(s), \\
\xi'&=x\,{\rm sh}(s)+\xi\,{\rm ch}(s),&\quad\xi'&=y\,{\rm sh}(s)+\xi\,{\rm
ch}(s),&\quad\xi' &=z\,{\rm sh}(s)+\xi\,{\rm ch}(s).
\end{alignat*}
Let's concentrate for a while on the flow in the plane
$(x,\xi)$.  It is obtained by solving the equations
\[
\frac{dx}{ds}=\xi,\quad \frac{d\xi}{ds}=x.
\]
Let $x_{s=0}=x_\circ,\ \ \xi_{s=0}=\xi_\circ$. Then the solution is
\[
x=x_\circ {\rm ch}(s)+\xi_\circ {\rm sh}(s)=
\frac{x_\circ +{\rm th}(s)\xi_\circ}{\sqrt{1-{\rm th}^2(s)}}=
\frac{x_\circ +\beta ct_\circ}{\sqrt{1-\beta^2}},
\]
\[
\xi=x_\circ {\rm sh}(s)+\xi_\circ {\rm ch}(s)=
\frac{x_\circ {\rm th}(s)+ct_\circ}{\sqrt{1-{\rm th}^2(s)}}=
\frac{x_\circ\beta+ct_\circ}{\sqrt{1-\beta^2}},
\]
where $\beta^2={\rm th}^2(s)\le 1$, and $s$ is fixed. The standard physical
interpretation of these relations is that the frame $(x_\circ, \xi_\circ)$
moves with respect to the frame $(x,\xi)$ along the common axis $x\equiv
x_\circ$ with the velocity $v=\beta c$, and since
$|\beta|\le 1$ then $|v|\le c$.
It is important to have in mind that this interpretation requires that $c$
has the same value in all such frames. This special invariance property of
Maxwell equations, i.e. of the complex structure $\Phi$, brought to life the
Poincar\'e-Einstein relativity principle.
\vskip 0.4cm
4. The dilatation vector field generates the flow:
\[
x^{\mu'}=ax^\mu, \ \ a=exp(s)=\text{const}.
\]
\vskip 0.4cm
5. The special conformal (with respect to the bilinear form $h^2$) vector
fields generate the nonlinear flows
\begin{alignat*}{4}
x'& =\frac{x+d^1(-x^2-y^2-z^2+\xi^2)}{1+2(-xd^1-yd^2-zd^3+\xi d^4)+
(-x^2-y^2-z^2+\xi^2)\left[-(d^1)^2-(d^2)^2-(d^3)^2+(d^4)^2\right]}&, \\
y'& =\frac{y+d^2(-x^2-y^2-z^2+\xi^2)}{1+2(-xd^1-yd^2-zd^3+\xi d^4)+
(-x^2-y^2-z^2+\xi^2)\left[-(d^1)^2-(d^2)^2-(d^3)^2+(d^4)^2\right]}&, \\
z'& =\frac{z+d^3(-x^2-y^2-z^2+\xi^2)}{1+2(-xd^1-yd^2-zd^3+\xi d^4)+
(-x^2-y^2-z^2+\xi^2)\left[-(d^1)^2-(d^2)^2-(d^3)^2+(d^4)^2\right]}&, \\
\xi'& =\frac{\xi+d^4(-x^2-y^2-z^2+\xi^2)}{1+2(-xd^1-yd^2-zd^3+\xi d^4)+
(-x^2-y^2-z^2+\xi^2)\left[-(d^1)^2-(d^2)^2-(d^3)^2+(d^4)^2\right]},
\end{alignat*}
where $(d^1,\dots,d^4)$ are the four constants-parameters of the
special conformal transformations. Note that these transformations may be
considered as coordinate transformations only if the corresponding
denominators are different from zero.

The above four relations may be written together in the following way
\[
x^{\mu'}=\frac{x^\mu+d^\mu(h^1_{\alpha\beta}x^\alpha x^\beta)}
{1+2h^1_{\alpha\beta}d^\alpha x^\beta +
(h^1_{\alpha\beta}x^\alpha x^\beta)(h^1_{\mu\nu}d^\mu d^\nu)}.
\]

These symmetry considerations show undoubtedly the above mentioned analogy
with the symplectic mechanics: the canonical $(q,p)$-transformations defined
as symmetries of the symplectic 2-form on $T^*(\mathbb{R}^3)$ determine
the symmetries of the hamilton equations; in the same way, the
transformations of $\mathbb{R}^4$, defined as (or generated by) symmetries of
the complex structure $\Phi$, determine the symmetries of Maxwell equations.

\subsection{Extended Electrodynamics}
Extended Electrodynamics (EED) was brought to life in pursue of a
theoretical ability to describe mathematically {\it time-stable finite field
configurations} having the basic features of finite field
objects:  to carry finite energy-momentum and spin-momentum, to propagate as
a whole along a given spatial direction with the velocity of light, to have
polarization properties, etc. From formal point of view EED extends the
Classical electrodynamics (CED) vacuum equations \linebreak $\mathbf{d}F=0,\
\mathbf{d}*F=0 $ to nonlinear vacuum equations. In standard Minkowski space
terms this extension looks as follows:
\begin{equation}
F\wedge *\mathbf{d}F=0,\quad (*F)\wedge
*\mathbf{d}*F=0, \quad F\wedge *\mathbf{d}*F+(*F)\wedge*\mathbf{d}F=0.    
\end{equation}
Note that instead of the Hodge $*$ we may introduce the operator
$\circledast$. In terms of the coderivative $\delta$ we have
\[
\delta F\wedge*F=0,\quad (\delta *F)\wedge F=0, \quad \delta F\wedge F-(\delta
*F)\wedge(*F)=0.
\]

In order to make use of the introduced through $\Phi$ in the previous
subsection operators we first recall [10] that the insertion operator $i(x)$
may be extended to an insertion operator $i(t):\alpha\rightarrow i(t)\alpha$,
where $t$ is a $q$-vector, $\alpha$ is a $p$-form, $i(t)\alpha$ is a
$(p-q)$-form, and $q\le p$.  For the decomposable case $t=x_1\wedge
x_2\wedge\dots\wedge x_q$ this extension is defined by
\[
i(t)\alpha=i(x_q)_\circ\dots _\circ i(x_1) \alpha,
\]
and in the nondecomposable case it is extended by linearity. In components
we obtain
\[
\left(i(t)\alpha\right)_{i_1\dots i_{p-q}}=
t^{k_1\dots k_q}\alpha_{k_1\dots k_q i_1\dots i_{p-q}},\quad
k_1< k_2<\dots<k_q, \quad i_1<\dots<i_{p-q},
\]
where summation over $k_1,\dots,k_q$ is understood.

In components equations (27) are given by
\[
F^{\mu\nu}(\mathbf{d}F)_{\mu\nu\sigma}=0,\quad
(*F)^{\mu\nu}(\mathbf{d}*F)_{\mu\nu\sigma}=0,\quad
F^{\mu\nu}(\mathbf{d}*F)_{\mu\nu\sigma}+
(*F)^{\mu\nu}(\mathbf{d}F)_{\mu\nu\sigma}=0,\ \mu<\nu.
\]
Hence, we can make use
of the operator $\mathfrak{D}$ to get $\mathfrak{D}F$ and
$\mathfrak{D}\Phi F=-\mathfrak{P}_\circ\Phi_\circ\Phi F=\mathfrak{P}F$, and
then we form the corresponding products.  So, the EED vacuum equations
(27) look as follows in these terms
\begin{equation}
i(\mathfrak{D}F)\mathbf{d}F=0,\quad
i(\mathfrak{P}F)\mathbf{d}(\Phi F)=0,\quad               
i(\mathfrak{D}F)\mathbf{d}(\Phi F)+
i(\mathfrak{P}F)\mathbf{d}F=0.
\end{equation}
Hence, EED vacuum equations also do {\it not} need and do {\it not} use
pseudoeuclidean metric.

We are going to give one more formulation of the EED vacuum equations, and to
this end we first extend the Lie derivative operator. Usually the Lie
derivative is defined with respect to a given (but arbitrary) vector field
and, considered in the frame of differential forms, it is given by $L_X
\alpha=i(X)\mathbf{d}\alpha +\mathbf{d}i(X)\alpha$, where $X$ is a vector
field and $\alpha$ is a $p$-form. Thus, $L_X \alpha$ describes how $\alpha$
changes along the trajectories of $X$. A natural question arises: is it
possible to define a "generalized Lie derivative", which more or less would
describe the {\it simultaneous} change of $\alpha$ along several
linearly independent vector fields, i.e along some distribution on the base
manifold?  In order to answer (although in some extent) positively to this
question we make the following construction.

Let $X_1,X_2,\dots,X_q$ be vector fields on an $n$-manifold $M$.
Then we have the $q$-vector \linebreak
$T=X_1\wedge X_2\wedge\dots\wedge X_q$.
Let $\alpha$ be a $p$-form on $M$ and $q\le p$. Then the generalized
Lie derivative $\mathfrak{L}_T\alpha$ of $\alpha$ along the $q$-tensor $T$ is
defined by
\begin{equation}
\mathfrak{L}_T\alpha=i(T)\mathbf{d}\alpha+\mathbf{d}i(T)\alpha.
\end{equation}
As in the usual case we always have commutation with $\mathbf{d}$:
$$
\mathfrak{L}_T\,\mathbf{d}=\mathbf{d}\,\mathfrak{L}_T.
$$
Of course, the above definition
may be immediately extended by linearity and used for any two $q$-vector and
$p$-form if $q\le p$, and, if $q>p$, it seems naturally to
put $\mathfrak{L}_T\alpha\equiv 0$.
If the vector fields $X_1,\dots,X_q$ define an integrable
distribution, and the Pfaff 1-forms
$\alpha^1,\alpha^2,\dots,\alpha^p$ define the corresponding
intagrable co-distribution,
$\langle\alpha^i,X_j\rangle=0, \ q+p=n$, then
\[
\mathfrak{L}_T(\alpha^1\wedge\alpha^2\wedge\dots\wedge \alpha^p)=0.
\]
Now the EED vacuum equations (28) are given by
\begin{gather*}
\mathfrak{L}_{\mathfrak{D}F}F=0,\quad
\mathfrak{L}_{\mathfrak{P}F}\Phi F=0,\quad
\mathfrak{L}_{\mathfrak{D}F}\Phi F+\mathfrak{L}_{\mathfrak{P}F}F=0,\\
\quad i(\mathfrak{D}F)F=0,\quad i(\mathfrak{D}F)\Phi F=0,
\end{gather*}
where the last two equations require zero values of the two well known
invariants $I_1$ and $I_2$.

\section{Conclusion}
Here we are going to mention those points of the paper which from our
point of view seem most important.

The duality properties of the solutions to Maxwell equations reveal the
internal structure of the field as having {\it two vector components}, which
are

-differentially interrelated (through the equations), but

-algebraically distinguished.

From the point of view advocated in this paper, an adequate understanding of
these duality properties of the solutions to Maxwell equations {\it requires
explicitly introduced complex structure} in the equations.  This brought us
to make use of $\mathbb{R}^2$-valued differential forms, $\omega$ and
$\Omega$, as mathematical model objects of the electromagnetic field. This is
the {\it first important step} towards further analysis.

In the traditional 3-dimensional formulation of the theory it is important to
introduce the canonical complex structure $\mathcal{I}$ of $\mathbb{R}^2$ in
the equations in order to be able to represent {\it any} solution in {\it
any} basis of $\mathbb{R}^2$, the presence of $\mathcal{I}$ makes it
possible.  Having this at hand we showed that the unimodular symmetries of
$\mathcal{I}$, i.e.  the 2-dimensional rotations, are in a close connection
with the conservative properties of the energy-momentum.

Much more interesting and fruitful turned out to be the 4-dimensional
formulation. Getting together the first couple (1) of Maxwell equations into
one relation through the appropriately constructed 2-form $F$,
$\mathbf{d}F=0$, made somewhere in the beginning of 20th century, we consider
as a great achievement in field theory, because no new or extra objects are
needed to formulate this couple of equations. The exterior derivative
$\mathbf{d}$ commutes with every transformation $f:
\mathbb{R}^4\rightarrow\mathbb{R}^4$, so, the symmetries of the whole Maxwell
system of equations shall be determined by the second couple (2) of
equations. The particular symmetry
$(\mathbf{E},\mathbf{B})\rightarrow(\mathbf{-B},\mathbf{E})$ interchanges the
two couples of equations, so, the second couple (2) of equations should also
be possible to be cast into the form of $\mathbf{d}F'=0$. The explicit form
of $F'$ is determined entirely by the equations: they require and make use of
the linear map $\Phi$, as given in the coordinate basis of
$\Lambda^2(T^*\mathbb{R}^4)$ by the matrix (19), and satisfying the condition
$\Phi_\circ\Phi=-id_{\Lambda^2(T^*\mathbb{R}^4)}$, which is the {\it defining
condition} for a linear map to be a {\it complex structure}. Hence, the
{\it only} external structure used by Maxwell
equations in their 4-dimensional formulation, is the complex structure
$\Phi$, {\it no pseudoeuclidean metric is needed}. Therefore, all symmetries
of these equations coming from transformations $f$ of the base manifold
$\mathbb{R}^4$ have to be symmetries of $\Phi$. These symmetries $f$ of
$\Phi$ include translations, spatial rotations, space-time (Lorentz)
rotations, dilatations, and (nonlinear space-time) special conformal
transformations. So, {\it the conformal transformations of the Minkowski
metric can be defined as those, leaving the complex structure $\Phi$
invariant}

 The physical interpretation of a Lorentz rotation as a
"change of an inertial frame with another inertial frame" requires the {\it
same} value of the constant $c$ in the time coordinate $\xi=ct$ with respect
to the various inertial frames.

We showed further that, combining appropriately $\Phi$ with the
naturally existing Poincar\'e isomorphism $\mathfrak{P}$, makes possible to
produce all pseudometric structures $h^p$ needed in the theory. Hence, all
these pseudometric structures are {\it secondary} objects, they are naturally
to be used in the theory but this is {\it not necessary}, and, therefore,
they should {\it not} be preliminary introduced and considered as necessary
(as it is usually done in many textbooks). The Poincar\'e-Einstein relativity
principle privileges only a {\it part} of the known symmetries of $\Phi$ and
says {\it nothing} about the rest part of the symmetries of $\Phi$.

It was further shown that Extended Electrodynamics {\it also} does {\it not}
need pseudoeuclidean metric structures, although it works with {\it
nonlinear} vacuum equations. Moreover, EED suggested a
natural generalization of the Lie derivative, giving how a $p$-form changes
along a given $q$-vector, i.e. along a $q$-distribution on the base manifold.
This generalization was used to give a new form if the EED vacuum equations,
and this last form recalls very much "symmetry conditions".

Finally we note that, having at hand the bilinear forms $h^p$, the
4-dimensional formulation of CED and EED in presence of external fields
(charges, currents, media) is straightforward, no special efforts are needed.

In conclusion we shall state once again our view, based on the above
considerations:

{\bf The only external structure required by CED and EED
in their 4-dimensional formulations on $\mathbb{R}^4$ is the complex
structure $\Phi$.  All pseudometric structures arise as secondary objects and
may be used when needed, but they should not be considered as
beforehand necessary structures in electrodynamics}.

\vskip 1cm
{\bf REFERENCES}
\vskip 0.5cm
1. {\bf Donev, S., Tashkova, M.},Annales de la Fondation Louis de Broglie, 23,
No.2 (1998),89-97;LANL e-print: patt-sol/9709006

2. {\bf Donev, S}, LANL e-print: hep-th/0006208

3. {\bf Heaviside, O.}, "Electromagnetic Theory" (London, 1893); Electrical
Papers, London, {\bf 1} (1892); Phil.Trans.Roy.Soc., 183A, 423 (1893)

4. {\bf Larmor, J.}, Collected Papers. London, 1928

5. "The Dirac Monopole", (Mir, Moscow, 1970)

6. {\bf Strajev, V.}, {\bf Tomilchik, L.}, "Electrodynamics with Magnetic
Charge", (Nauka i Tehnika, Minsk, 1975)

7. {\bf Gibbons, G.}, {\bf Rasheed, D.}, hep-th/9506035

8. {\bf Chruscinski, D.}, hep-th/9906227

9. {\bf Chan Hong-Mo}, hep-th/9503106; hep-th/9512173

10. {\bf Greub, W.}, "Multilinear Algebra", Springer-Verlag, 1967

\end{document}